\documentclass{iau}

\usepackage{amsmath}
\usepackage{natbib,aas_jrnl_names}
\usepackage{graphicx}
\usepackage{multirow}

\newcommand{\Vm}{$\mathrm{V_{mean}}$}

\newcommand{\Ms}{$\mathrm{M_\odot}$}

\begin{document}

\aopheadtitle{Proceedings of the IAU Symposium}
\editors{N. Gopalswamy,  O. Malandraki, A. Vidotto \&  W. Manchester, eds.}

\lefttitle{Mohan, A., et al.}
\righttitle{Solar-stellar CMEs and type-IV bursts}

\jnlPage{1}{7}
\jnlDoiYr{2024}
\doival{10.1017/xxxxx}
\volno{388}
\pubYr{2024}
\journaltitle{Solar and Stellar Coronal Mass Ejections}

\title{CME-associated type-IV radio bursts: The solar paradigm and the unique case of AD Leo.}
\author{Atul Mohan$^{1,2}$, Nat Gopalswamy$^1$, Surajit Mondal$^3$, Anshu Kumari$^{1}$, Sindhuja G.$^{4}$}
\affiliation{{}$^1$NASA Goddard Space Flight Center, 8800 Greenbelt Road Greenbelt, MD, 20771, USA}
\affiliation{{}$^2$The Catholic University of America, 620 Michigan Avenue, N.E. Washington, DC 20064, USA}
\affiliation{{}$^3$Center for Solar-Terrestrial Research, New Jersey Institute of Technology, NJ 07102-1982, USA}
\affiliation{{}$^4$Bangalore University, Mysore Rd, Jnana Bharathi, Bengaluru, Karnataka, IN}

\begin{abstract}
The type-IV bursts, associated with coronal mass ejections (CMEs), occasionally
extend to the decameter-hectrometric (DH) range. We present a comprehensive catalog of simultaneous multi-vantage point observations of DH type-IV bursts by Wind and STEREO spacecraft since 2006. 73\% of the bursts are associated with fast
($>$900\,km\,s$^{-1}$) and wide ($>60^\circ$) CMEs, which are mostly geoeffective halo CMEs. Also, we find that the bursts are best observed by the spacecraft located within $|60^\circ|$ line of sight (LOS), highlighting the importance of LOS towards active latitudes while choosing target stars for a type-IV search campaign. In young active M dwarfs, CME-associated bursts have remained elusive despite many monitoring campaigns. We present the first detection of long-duration type-III, type-IV, and type-V bursts during an active event in AD Leo (M3.5V; 0.4\Ms). The observed burst characteristics support a multipole model over a solar-like active region magnetic field profile on the star.


\end{abstract}


\maketitle

\section{Introduction}
Coronal mass ejections (CMEs) often produce radio bursts associated with the propagating shock and the reconnection event~\citep{Mclean85_book}.
These include the type-II, complex long duration type-III and type-IV bursts, each of which have characteristic emission features in the dynamic spectrum (DS)~\citep{Gopal11_PREconf}.
The type-II bursts are associated with coronal and interplanetary (IP) shocks, while complex type-IIIs with a duration of the order of an hour are signatures of accelerated electrons propagating along open field lines~\citep{2000GeoRL..27.1427G}.
Meanwhile the type-IV burst which is the subject of this article, comes in two dynamic spectral variants: the moving type-IV which is a broadband emission with a drifting central frequency, and the stationary type-IV which is a broadband time-varying emission~\citep{takakura61_typeIVtypes,kundu63_DScharacterisationTypeIV}.
The moving type-IV burst is caused by flare-accelerated electrons trapped in a magnetised plasmoid moving with the CME~\citep{1971AuJPh..24..229S,schahl72_mtypeIV_model}, while the latter variant is produced by supra-thermal electrons in the post-flare loops~\citep{wild1970}.
Due to its close association with various aspects of the CME, type-IV bursts are highly sought after in stellar CME search campaigns.
\section{The solar paradigm}
Like all radio burst types, the type-IV bursts were also identified in the sun and studied extensively over decades by contrasting with multi-waveband imaging and non-imaging data in the radio to the $\gamma$-ray band.
Due to the availability of regular ground-based radio dynamic spectral observations since 1950s~\citep{wild50_bursttypes}, several statistical studies on meter waveband bursts exist. 
Using 15 years of data, \cite{Cane88_typeIV_stats_XrayflareCorr} showed that metric type-IVs are mostly associated with strong long-duration soft X-ray (SXR) flares with a median flare strength of $\sim$ 3$\times$10$^-5$\,W\,m$^{-2}$ and a median duration of $\sim$1\,h.

The decameter-hectometric (DH; 1 -14\,MHz) type-IV bursts are rare continuations of the metric bursts, with 100\% association to white-light CMEs~\citep{Hillaris16_typeIV,Atul24_DHtypeIVcatalog}. These events could be explored only after the advent of space-based radio observations with the launch of Wind and later the STEREO spacecraft.
The CMEs and flares linked to DH type-IVs are much stronger than those associated with the type-IVs that terminate within the metric band. About 89\% of the DH type-IVs are associated with M to X class (10$^{-5}$ - 10$^{-4}$\,W\,m$^{-2}$)~\citep{Hillaris16_typeIV}.
\cite{Miteva17_SEP-radburst_link} showed that of all the radio bursts, DH type-IV bursts show the highest association with solar energetic particle (SEP) events. All these properties make DH type-IV bursts highly relevant in space weather studies. 
Besides, the distribution of type-IV source longitudes are found to prefer $|60^\circ|$ longitude~\citep{Cane88_typeIV_stats_XrayflareCorr,gopal16_typeIVdirectivity}. There has been a debate on this being caused by either an inherent emission beaming (directivity) expected from the coherent plasma emission mechanism~\citep{gopal16_typeIVdirectivity}, or purely a propagation effect by which dense plasma structures along the line of sight (LOS) occult the emission~\citep{Pohjo20_typeIVdirective,Nasrin18_DHtypeIV_occult_streamerCMEshock}.
The effect of directivity can be crucial in stellar radio burst searches because choosing a candidate star with a high chance of type-IV detection depends on the LOS towards its active latitudes. 
However, the inferences on the LOS effect are based on either a few event studies with simultaneous multi-spacecraft data, or statistics on single spacecraft data. There have been no studies on a statistically significant sample of events with simultaneous multi-vantage point observations from multiple spacecraft, to systematically explore the LOS effect. 
Firstly, we need a comprehensive catalog of DH type-IV bursts observed the different spacecraft over the last couple of solar cycles.
%
\subsection{The multi-vantage point DH type-IV catalog}
The DH dynamic spectra from Wind (since 1996), STEREO-A and STEREO-B spacecraft (since 2006) form the data source.
To search for type-IV events, three event lists were used; the LASCO CME catalog~\citep{Yashiro04_LASCOCME_Catalog}, a metric type-IV event list based on the daily reports from Space Weather Prediction Center (SWPC), and the DH type-II catalog~\citep{Gopal19_DHtypeII_catalog}.
Since previous studies show a 100\% association of the DH type-IV bursts with CMEs and since they are extensions of metric type-IVs, the union of the first two event lists should cover all DH type-IV events. 
The DH type-II list gives additional intervals to search in the DH band, in case some events were missed out due to data gaps.
For each date-time period in the search list, the DS from each spacecraft was examined. In the case of a potential type-IV detection, the corresponding metric DS was searched for to confirm the signature. All dubious events were removed.
\begin{figure}[t]
    \centering
  \includegraphics[width=0.82\textwidth, height=0.36\textheight]{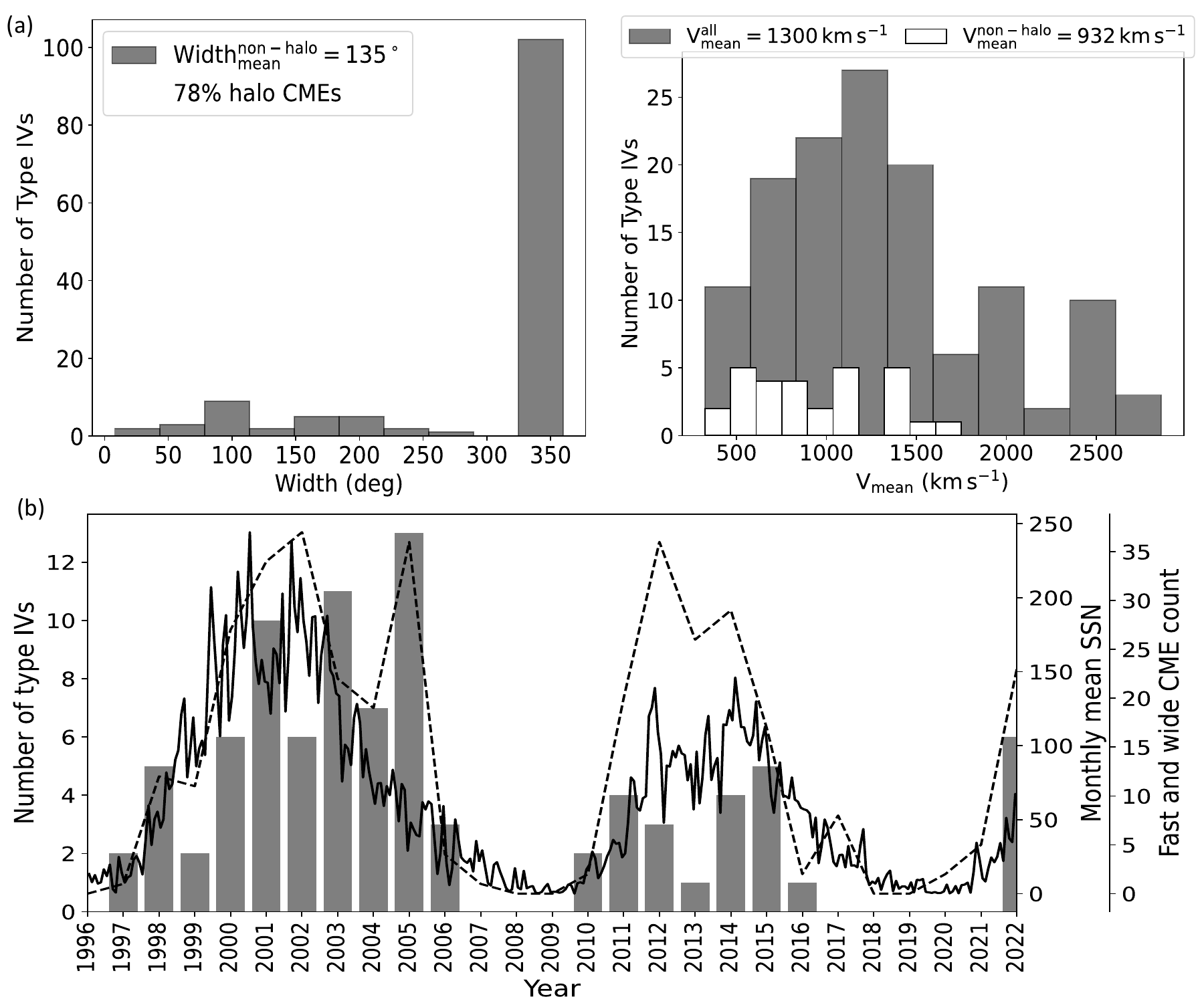}
  \vspace{-0.3cm}
  \caption{(a): Properties of CMEs associated with the DH type-IV bursts. (b): Solar cycle variation in the number of DH type-IV bursts (histogram), fast and wide CMEs (dotted), and mean sunspot number (solid).}
  \label{fig1:stats}
\end{figure}
\begin{figure}[]
\centering
  \includegraphics[width=0.94\textwidth, height=0.35\textheight]{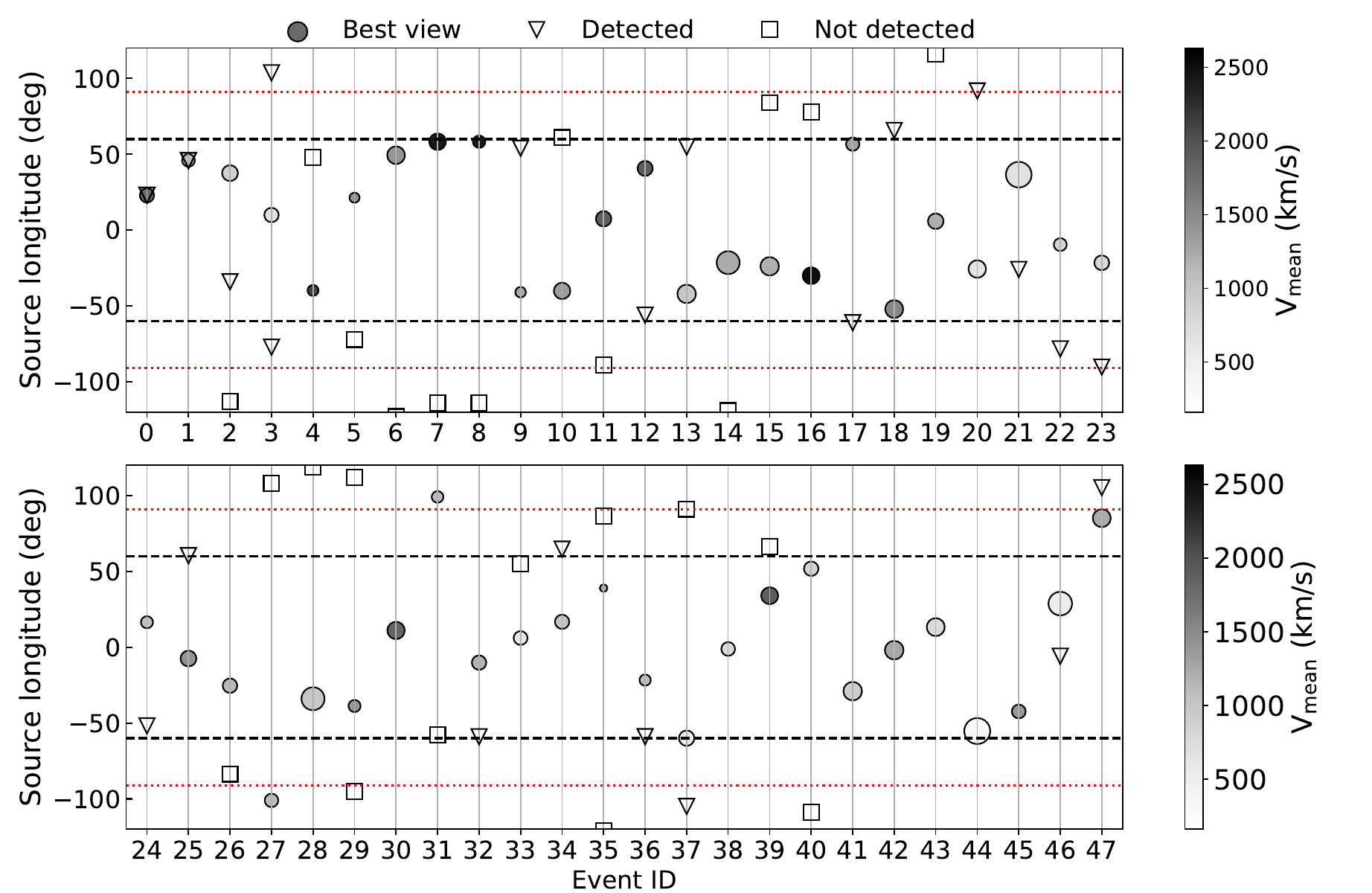}
  \vspace{-0.3cm}
  \caption{Events with good data from at least 2 spacecraft. Event ID simply labels each event. For each ID, the markers along the Y axis show the longitudes of the active region (AR) as seen by different spacecraft, providing the various lines of sight. Circle shows the LOS that gave the best view of the type-IV, its size denotes the burst duration and the color shows \Vm. Triangles mark the other detected longitudes and squares show non-detections. Horizontal lines mark $\pm$60$^{\circ}$ and $\pm$90$^{\circ}$ longitudes. Longitudes $>|90^\circ|$ pick out ARs that hosted high rising post-flare loops with type-IV emission visible beyond the limb.}
  \label{fig2:LOS}
\end{figure}

The final catalog has 139 DH type-IV bursts with a mean duration of $\sim$ 83\,min and extending down to $\sim$7.8\,MHz on average. The full catalog can be accessed online\footnote{\href{https://cdaw.gsfc.nasa.gov/CME\_list/radio/type4/}{https://cdaw.gsfc.nasa.gov/CME\_list/radio/type4/}}
All events had a white-light CME association in contrast to the type-IV bursts confined within the metric band~\citep{Anshu21_C24typIV_CME_corr}. 73\% of the DH type-IVs are linked to a fast ($>$900\,km\,s$^{-1}$) and wide ($>60^\circ$) CME, and $\sim$78\% to halo CMEs~\citep{howard82_haloCMEdiscovery,gopal07_haloCMEstats} which are a class of relatively fast CMEs (mean speed, \Vm$\sim$1050\,km\,s$^{-1}$) with nearly 360$^\circ$ angular width compared to regular CMEs. Figure~\ref{fig1:stats}a shows the properties of CMEs linked to DH type-IV bursts. With an average \Vm $\sim$1300\,km\,s$^{-1}$, even the non-halo CMEs linked to DH type-IVs are fast CMEs. 
Figure~\ref{fig1:stats}b shows the strong correlation between the occurrence of DH type-IV bursts and fast-wide CMEs over two solar cycles.

To explore the effect of LOS on the observed type-IV DS, we closely examine the 48 bursts which had good quality observations from at least two spacecraft simultaneously. Figure~\ref{fig2:LOS} shows the different bursts numbered by IDs along with the observed source longitudes or equivalently AR LOS in various spacecraft. The best-view longitude of an event is the LOS that observed the brightest burst with the largest frequency-time extent, out of all recorded simultaneous multi-vantage point data. A clear concentration of best-view longitudes within $|60^\circ|$ is seen.
Except for Event 31, in cases where the best-view longitude was $>|60^\circ|$, there was no spacecraft observing the AR within the $|60^\circ|$ LOS.
During Event 31, dense plasma structures of the neighboring AR caused significant LOS occultation to the spacecraft viewing from within $<|60^\circ|$ LOS.
Hence, we note that in the absence of strongly occulting structures along the LOS, the spacecraft observing an AR within $|60^\circ|$ LOS or source longitude range always records the best-view of the burst. In general smaller the magnitude of the LOS angle, better the view as evidenced by the triangle (and square) markers that blanket the circles in Fig.~\ref{fig2:LOS}.
For more details, we refer the reader to \cite{Atul24_DHtypeIVcatalog}.

\section{The curious case of young active stars: Lessons from AD Leo}
The cool main sequence stars, particularly the M dwarfs (dMs) within 0.6 - 0.1\,\Ms\ mass range are known to show high flaring rates across the electromagnetic spectrum~\citep[e.g.][]{Lacy76_Flare_stats_Mstars,bastian90_flarestarsRev,dal2020flare}. 
The dM masses are spread across the $0.4 - 0.3$\Ms\ band below which the stellar interior becomes fully convective, unlike a sun-like star that has an inner radiation zone~\citep{chabrier00_Stellainterior}.
In general, stellar atmospheres evolve with age as stars migrate from a fast-rotating active population (`C' branch) to a slow-rotating less active `I' branch as they lose their angular momentum over time~\citep{Barnes03_Rot_Vs_age_Vs_Activity,brown14_CI_branchtheroy}.
The stars above 1\,Gyr, like our Sun, generally exist in the less active `I' branch
~\citep{Barnes03_Rot_Vs_age_Vs_Activity}.
Hence young ($<$1\,Gyr) dMs with mass $\lesssim$ 0.4\Ms\ are expected to be very different from the sun in their internal structure and atmospheric activity type.

Young dMs demonstrate some of the highest flaring rates with a high ratio of flare-to-quiescent luminosity in X-ray and spectral lines~\citep[e.g.,][]{Noyes84_RHK,2003A&A...397..147P,Vidotto14_B_Vs_age_n_rot,Newton17_Rha_dMs,huiqin19_Rflare_vs_Ro_keplerdata,feinstein20_flarestats_youngstars}. But, despite the high activity, there has been no radio signatures of CMEs in the form of either a type-II or type-IV bursts in these stars, despite several hours to day long monitoring by various groups for over a few decades~\citep{Osten08_ADLeoFinebursts,Villadsen14_First_detect_SLS_inRadio_VLA}. The lack of type-II bursts is attributed to the strong magnetic field strengths in the corona which suppresses the shock formation~\citep{alvarado22_CME_star}. However, the missing stationary type-IVs still remained a puzzle, as they do not need a shock but just a population of flare-accelerated electron beams within the post-flare loop.
A possible reason for missing out on type-IV detection could be the effect of LOS or emission directivity.
In fact there has been only one reported type-IV detection so far in a star other than the Sun~\citep{Zic20_typeIV_ProximaCen}.
\begin{figure}[t]
    \centering
  \includegraphics[width=0.98\textwidth, height=0.2\textheight]{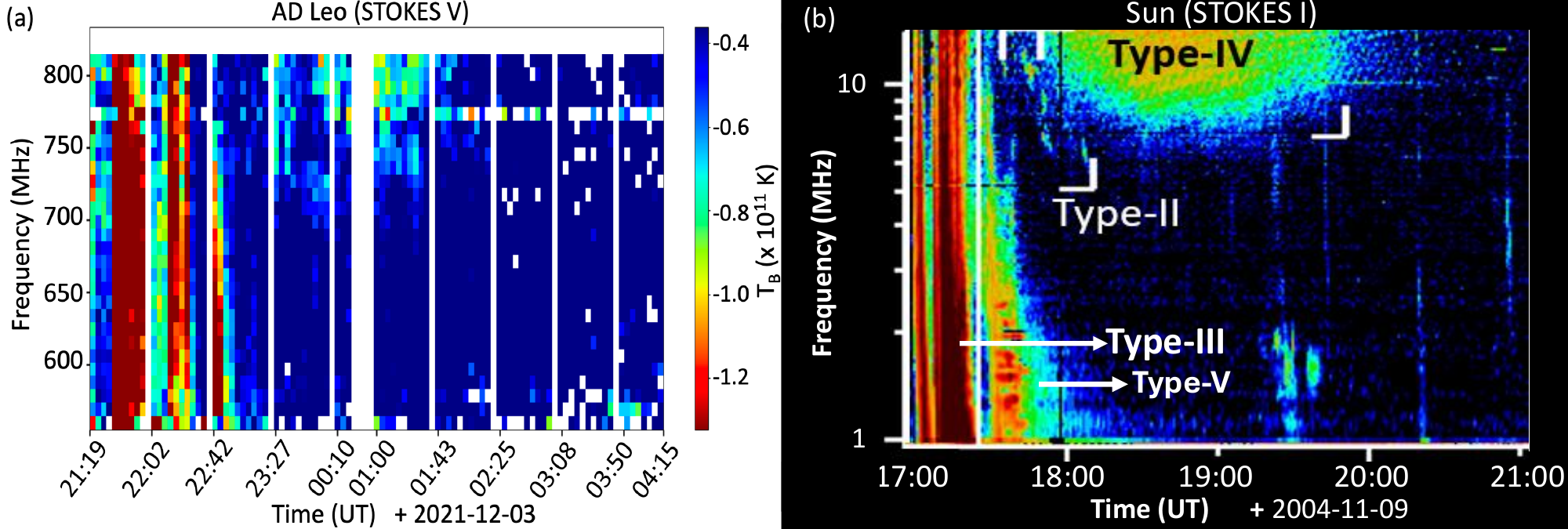}
  \vspace{-0.1cm}
  \caption{(a): AD Leo DS showing long duration type-III and type-IV bursts. 
  A faint type-V signature is also noted. (b) A sample solar DS from DH type-IV catalog during a CME, with the burst types marked.}
  \label{fig:typeIVADLeo}
\end{figure}

With an age of $\sim$ 250\,Myr old, rotation period of 2.2\,d and a mass around 0.4\Ms, AD Leo (M3.5V) is a non-solar type star in terms of atmospheric activity type and internal structure.
Zeemann Doppler Imaging (ZDI) studies show that the stellar rotation axis has an inclination of $\sim$20$^\circ$ along the Earth LOS~\citep{2010MNRAS.407.2269M}, making the south pole and the surrounding latitudes directly visible. In active young dMs, observations suggest that the flaring ARs are present in poleward latitudes than being focused within $|45^\circ|$ like in the Sun~\citep{Ilin21_highlat_activeReg}. 
Besides, recent ZDI data show signs of a magnetic structure transition in AD Leo since 2020, whereby the axisymmetry of the large scale field dropped to around 60\% from above 90\% in the last decade~\citep{Bellotti23_ZDIADLeo_Bevol}. The reduction in the global poloidal field could present an opportunity for the massive flares to drive eruptions with a higher success rate as the strength of the Alfv\'en barrier proposed to effectively block eruptions or at least damp their energy so that a shock does not develop, will now be weaker. 
Using upgraded Giant Metrewave Radio Telescope (uGMRT), we observed AD\,Leo from 3 December 2021, 21:17 to
4 December 2021, 04:17 UT. We report the discovery of long-duration type-III bursts followed by a type-IV radio burst on a young dM. 
Besides being the first type-IV in a young dM, this is the first detection of a combination of CME-associated bursts in a star other than Sun, during an active event.
Figure~\ref{fig:typeIVADLeo} highlights the correspondence of the AD Leo DS with a solar CME-associated DS. 

The observed long-duration type-III, type-IV and type-V (type-III-like feature with a gradually varying duration as a function of frequency) signatures are likely powered by an electron cyclotron maser emission mechanism, as opposed to plasma emission processes~\citep{ginzburg1958, melrose1970} seen in the solar case. This can be attributed to the stronger magnetic fields on AD Leo. \cite{atul24_ADLeotypeIV} compared different magnetic field profiles that are compatible with the observed burst emission properties and concluded that a solar-like AR magnetic profile cannot explain AD Leo bursts. Instead, a simple multipole expansion model with the relative strengths in different poles adopted from recent ZDI results~\citep{Bellotti23_ZDIADLeo_Bevol}, agrees better with the data.
\section{Conclusion}
The type-IV bursts are closely linked to solar CMEs, with the decameter-hectometric (DH) bursts being 100\% associated with CMEs that cause interplanetary shocks. The stationary type-IV bursts are signatures of accelerated electron activity in the post-eruption arcades, while moving type-IV are produced by supra-thermal electron in moving plasmoids associated with the CME.
We present a comprehensive multi-vantage point catalog of decameter-hectmetric (DH) type-IV bursts. They typically extend down to 7.8\,MHz and have an average duration of 83\,min. These bursts are associated mostly with fast ($>$900\,km\,s$^{-1}$) and wide ($>$60$^\circ$) CMEs, with 78\% association to halo-CMEs making them highly geo-effective and space weather relevant.
The flux and morphology of the observed type-IV bursts in the dynamic spectrum (DS) are significantly affected by the line of sight (LOS). It is found that, unless there are dense plasma structures along the LOS, a spacecraft that observes the flaring active region (AR) within $|60^\circ|$ always record the burst with the highest extent in the frequency-time plane, out of all recorded DS from various vantage points. 
Based on this, we propose that choosing stars that provide direct LOS to active latitudes can better the odds for type-IV detection.

We present the first detection of CME-associated long-duration type-III and type-IV bursts in a young ($\sim$250\,Myr) active dM, AD Leo (M3.5V; 0.4\Ms). This is the fist case of detecting a combination of multiple CME-associated solar busts (long duration type-III, type-V and type-IV) producing a typical solar CME DS in a star other than the Sun.
The stellar rotation axis inclination of $\sim20^\circ$ along the LOS and the low axisymmetry fraction for the large-scale magnetic field over the recent years possibly contributed to the successful detection. The bursts are likely powered by an electron cyclotron maser (ECME) mechanism than the plasma emission processes that dominate in solar radio bursts. Besides, a multipole expansion model better fits the AR magnetic field profile than a solar-like profile given the type-IV characteristics. 
\bibliography{Sample}
\bibliographystyle{iau}
\end{document}